\documentclass[twocolumn,aps,floatfix,showpacs,prl]{revtex4}
\usepackage{amsmath}
\usepackage{graphicx}
\usepackage{psfrag}
\usepackage[T1]{fontenc}

\begin{document}

\title{Creation on demand of higher orbital states in a vibrating optical lattice}
\author{Tomasz Sowi\'nski}
\affiliation{
\mbox{$^1$Institute of Physics of the Polish Academy of Sciences, Al. Lotnik\'ow 32/46, 02-668 Warsaw, Poland}}
\date{\today}
\begin{abstract}
It is shown that the extended Hubbard Hamiltonian describing atoms confined in an optical lattice always contains commonly neglected terms which can significantly change the dynamical properties of the system. Particularly for bosonic systems, they can be exploited for creating orbital states on demand via the parametric resonance phenomenon. This indicates an additional application for optical lattices, namely the study and emulation of interactions between particles and lattice vibrations. 
\end{abstract}
\maketitle
The past decade of experiments on cold atom systems confined in optical lattices has brought huge progress in simulating different variations of the Hubbard model \cite{Zwerger,LewensteinRev}. It has become possible to experimentally mimic not only the simplest Bose-Hubbard model \cite{Zoller,Greiner}, but also to study systems with internal degrees of freedom \cite{Baranov,Lahaye}. Recent theoretical works \cite{Girvin,Kuklov,HemmerichT,Liu} and the first experiments considering higher bands in optical lattices \cite{Bloch,Hemmerich} have opened a new, promising area of ultra cold atom research -- orbital physics \cite{LewensteinLiu}. Typically, such systems are described with Hubbard-like Hamiltonians extended by additional inter-orbital interaction terms. In previous analyses, some of these additional terms, due to their non resonant character, have always been treated as unimportant and therefore omitted. In this letter, I give a simple counter-example that shows that in the case of oscillating optical lattices they can efficiently transfer atoms to higher bands in a fully controlled way. All the arguments presented here are given directly for bosons interacting via $\delta$-like interactions. Nevertheless, the central observation made here originates from the structure of any many-body Hamiltonian. Therefore, it can be easily adopted for other interparticle interactions as well as for fermions.

For simplicity, let me consider spinless (or polarized) bosons interacting via two body $\delta$-like contact interactions confined in a 2D optical lattice $V_{\mathtt{ext}}(\boldsymbol{r}) = q_x \sin^2(kx) + q_y \sin^2(ky) + \frac{m\omega_z^2}{2}z^2$,
where $k = 2\pi/\lambda$ is the wave vector of the laser field and $\omega_z$ is the frequency of the trapping harmonic potential in the $z$ direction. The optical lattices need not have the same depth in the $x$ and $y$ directions and they can be changed independently. The general Hamiltonian of this system written in second quantization formalism has the form ${\cal H} = \int\!\!\mathrm{d}^3\boldsymbol{r}\, \Psi^\dagger(\boldsymbol{r})\left[-\frac{\hbar^2}{2m}\nabla^2 + V_{\mathtt{ext}}(\boldsymbol{r})\right]\Psi(\boldsymbol{r})
+ \frac{g}{2}\int\!\!\mathrm{d}^3\boldsymbol{r}\,\Psi^\dagger(\boldsymbol{r})\Psi^\dagger(\boldsymbol{r})\Psi(\boldsymbol{r})\Psi(\boldsymbol{r})$,
where $\Psi(\boldsymbol{r})$ is a bosonic field operator and $g$ is the strength of the contact interactions. It is convenient to measure lengths in units of the laser wavelength $\lambda$, and all energies in units of the recoil energy $E_R = \frac{(2\pi\hbar)^2}{2m\lambda^2}$. The trapping potential in the $z$ direction is characterized by the dimensionless quantity $\kappa = \hbar \omega_z/2E_R$. The dimensionless coupling constant is $g = 16\pi^2 a_0/\lambda$, where $a_0$ is the s-wave scattering length. In typical experiments with $^{87}\mathrm{Rb}$ or $^{52}\mathrm{Cr}$ confined in an optical lattice, far from the Feshbach resonance, $g \sim 1$. To derive an extended Bose-Hubbard model describing this system, one expands the field operator in the ground and first excited Bloch bands as follows: $\Psi(\boldsymbol{r})\approx \sum_i \hat{a}_i\phi_i^0(\boldsymbol{r}) + \hat{b}_i\phi_i^x(\boldsymbol{r})+ \hat{c}_i\phi_i^y(\boldsymbol{r})$. The functions $\phi_i^0(\boldsymbol{r}) = {\cal X}_i^{0}(x){\cal Y}_i^{0}(y){\cal Z}(z)$, $\phi_i^x(\boldsymbol{r}) = {\cal X}_i^{1}(x){\cal Y}_i^{0}(y){\cal Z}(z)$, and $\phi_i^y(\boldsymbol{r}) = {\cal X}_i^{0}(x){\cal Y}_i^{1}(y){\cal Z}(z)$ are products of one-dimensional Wannier functions ${\cal X}_i^\alpha(x)$ (${\cal Y}_i^\alpha(y)$) from band $\alpha$ localized in site $i$, and the ground state of the harmonic oscillator in the $z$ direction ${\cal Z}(z) = (\kappa/\pi)^{1/4}\mathrm{exp}(-\kappa z^2/2)$. This decomposition is valid provided that $\kappa^2 \gg q_x/E_R$ and $\kappa^2 \gg q_y/E_R$. In such a case, the energy gap for excitations in the $z$ direction is much larger than the gaps in the lattice directions, and therefore particle dynamics is frozen in the $z$ direction. Bosonic operators $\hat{a}_i$, $\hat{b}_i$, and $\hat{c}_i$ annihilate particles at site $i$ in the $s$, $p_x$, and $p_y$ orbitals respectively. For convenience, I introduce particle number operators $\hat{n}^{(i)}_s = a^\dagger_i a_i$, $\hat{n}^{(i)}_x = b^\dagger_i b_i$, $\hat{n}^{(i)}_y = c^\dagger_i c_i$ as well as the dimensionless algebraic vector $\boldsymbol{Q}=(q_x/E_R,q_y/E_R,\kappa)$ which characterizes the geometry of the optical lattice. In the above approximation, the Bose-Hubbard Hamiltonian takes the form:
\begin{subequations} \label{Hamiltonian}
\begin{align} 
{\cal H} = \sum_i H_i&-\sum_{\{i\xrightarrow{\mathrm{x}}j\}} J^x_0 (a_i^\dagger a_j + c_i^\dagger c_j) + J^x_1 b_i^\dagger b_j \nonumber \\
&-\sum_{\{i\xrightarrow{\mathrm{y}}j\}} J^y_0 (a_i^\dagger a_j + b_i^\dagger b_j) + J^y_1 c_i^\dagger c_j,
\end{align}
where $J^d_\alpha$ is the standard one dimensional nearest neighbor hopping amplitude in the direction $d$ for band $\alpha$. The summation $\sum_{\{i\xrightarrow{\mathrm{d}}j\}}$ is understood as a summation over all sites $i$ and over all nearest neighbors $j$ of site $i$ in the direction $d$. The on-site Hamiltonian $H_i$ is a sum of the single particle energies and two-body interaction terms and has the form:
\begin{align}
H_i &= \sum_\sigma \left[E_\sigma \hat{n}_i^\sigma + \frac{U_{\sigma\sigma}}{2} \hat{n}_i^\sigma(\hat{n}_i^\sigma-1)\right] + \sum_{\sigma\neq\sigma'} U_{\sigma\sigma'} \hat{n}_i^\sigma\hat{n}_i^{\sigma'} \nonumber \\
&+ \left[\frac{U_{sx}}{2}\,\hat{a}_i^\dagger{}^2\hat{b}_i{}^2 + \frac{U_{sy}}{2}\,\hat{a}_i^\dagger{}^2\hat{c}_i{}^2 + \frac{U_{xy}}{2}\,\hat{b}_i^\dagger{}^2\hat{c}_i{}^2 \right] + h.c. \label{NotResTerms}
\end{align}
\end{subequations}
These summations run over orbital index $\sigma\in\{s,x,y\}$. The single particle energies $E_\sigma$ depend only on the lattice geometry while all the parameters $U$ depend additionally on the dimensionless coupling $g$. They can be calculated directly:
\begin{align}
E_\sigma(\boldsymbol{Q})&=\int\!\mathrm{d}^3\boldsymbol{r}\,{\phi_i^\sigma}(\boldsymbol{r})\left[-\frac{\hbar^2}{2m}\nabla^2 + V_{\mathtt{ext}}(\boldsymbol{r})\right]\phi_i^\sigma(\boldsymbol{r}), \nonumber \\
U_{\sigma\sigma'}(\boldsymbol{Q},g) &= g\! \int\!\mathrm{d}^3\boldsymbol{r}\left[\phi_i^\sigma(\boldsymbol{r})\phi_i^{\sigma'}(\boldsymbol{r})\right]^2.
\end{align}
\begin{figure} 
\includegraphics{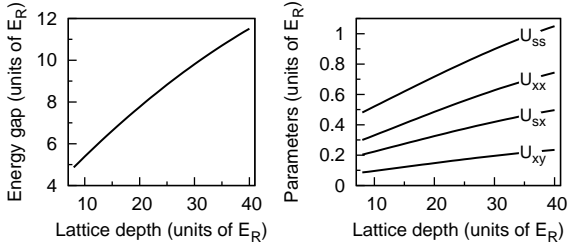}
\caption{Parameters of the Hamiltonian \eqref{Hamiltonian} as functions of lattice depth for the symmetric case $\boldsymbol{Q}=(q,q,8)$ and $g = 1$. Parameters originating in contact interactions are at least ten times smaller than the energy gap between the $s$ and $p$-bands.} \label{FigParam}
\end{figure}
To get more insight into the system, the values of these parameters for the example of a symmetric lattice $\boldsymbol{Q}=(q,q,8)$ and $g = 1$ are presented in Fig.\ref{FigParam}. For experimentally available systems, far from the Feschbach resonance, all contact energies $U$ are at least ten times smaller than single-particle excitation energies $\Delta E_\sigma = E_\sigma - E_s$ where $\sigma\in\{x,y\}$. Usually, one can neglect the two next to last terms in the Hamiltonian \eqref{NotResTerms}, if the contact interaction energy is a small correction to the gap energy between bands. These terms describe interaction processes which transfer two particles between bands and they are highly suppressed since they violate energy conservation. However, if one considers a scenario when the lattice parameters $\boldsymbol{Q}$ or the coupling constant $g$ vary in time, that argument is no longer valid since energy conservation simply does not hold in such a case. Nevertheless, in previous analyses \cite{Girvin,Kuklov,Liu} these non resonant terms were always neglected, even for fast-varying Hamiltonians \cite{Zurek,Schutzhold,Sengupta}. Under such a far-fetched approximation, particle numbers in each orbital are constant. In this paper we utilize these commonly neglected terms and propose a mechanism for creating orbital states {\it on demand}. To show that this scenario is realistic in present day experiments, numerical simulations are shown for $^{52}\mathrm{Cr}$ atoms confined in an optical lattice with $\lambda = 523\,\mathrm{nm}$. The contact interaction coupling is $g\approx 1.8$. 
\begin{figure}
\includegraphics{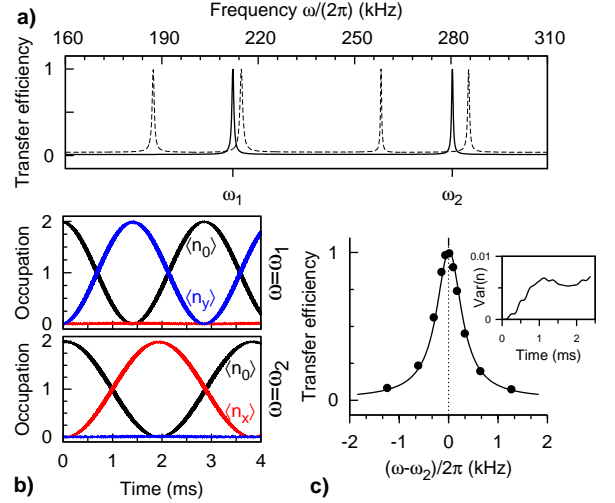}
\caption{Creation of the $p$-band states by a vibrating optical lattice with $\boldsymbol{Q}=(32+4\sin(\omega t),20,8)$. Plot (a) presents the transfer efficiency as a function of vibration frequency $\omega$. Two distinct peaks are visible. For these particular frequencies, $p$-band states become highly occupied. The dashed line comes from a generalized model that also takes into account $d$-band orbitals. Additional peaks correspond to resonant frequencies in which one of interacting  atoms can be promoted to the $d_x$ or $d_y$ band, respectively. Moreover, resonant frequencies $\omega_1$ and $\omega_2$ are shifted. Plot (b) presents the occupation of basis states as a function of time for the resonant frequencies. At frequency $\omega_1$ ($\omega_2$) the $p_y$ ($p_x$) orbital is populated. Plot (c) shows a comparison between the situation when tunneling is totally neglected (solid line) and when it is taken into account (filled circles). The inset shows the variance of the on-site number operator as a function of time for $\omega=\omega_2$. Detailed explanations are given in the text. } \label{resonanceFig}
\end{figure}

The starting point of the analysis is the many-body ground state of bosons confined in a static and deep optical lattice $\boldsymbol{Q}_0=(32,20,8)$ filled with two bosons per site on average. In such a case, tunneling is highly dominated by the on-site contact interactions, and the many body ground state is the Mott Insulator in the $s$-band. Hence the state of the system can be characterized quite well by a product of local ground states in independent lattice sites. Now let us study the situation when the lattice parameters change periodically in time in the following way: $\boldsymbol{Q}(t) = \boldsymbol{Q}_0 + \delta\boldsymbol{Q}(t)$. In the simplest case, the lattice parameters oscillate with some amplitude $A$ and frequency $\omega$ in one spatial direction only, i.e. $\delta\boldsymbol{Q}(t)=(A,0,0)\sin(\omega t)$. Since the lattice depth is large for our choice of $\boldsymbol{Q}$ the tunneling processes are very slow. Therefore it will be assumed that the dynamics in each site is independently governed by the single-site Hamiltonian $H_i$ and the number of particles in each site is conserved. The correctness of this assumption was verified with dynamical many-body calculations and is discussed below. At the initial moment, two particles occupy the $s$-band state. Therefore, due to the structure of the Hamiltonian \eqref{NotResTerms}, the entire dynamics takes place in the subspace spanned by three states: $|200\rangle=\frac{1}{\sqrt{2}}a^\dagger{}^2|\mathrm{vac}\rangle$, $|020\rangle=\frac{1}{\sqrt{2}}b^\dagger{}^2|\mathrm{vac}\rangle$, and $|002\rangle=\frac{1}{\sqrt{2}}c^\dagger{}^2|\mathrm{vac}\rangle$. In this subspace, the Hamiltonian has a simple matrix form\begin{align} \label{HamMatrix}
\hat{H}(\boldsymbol{Q}(t)) = \left( 
  \begin{matrix}
    2E_s+U_{ss} & U_{sx} & U_{sy} \\
    U_{sx} & 2E_x + U_{xx} & U_{xy} \\
    U_{sy} & U_{xy} & 2E_y + U_{yy}
  \end{matrix}\right).
\end{align}
All the parameters of this matrix depend on time through the time dependence of the lattice shape $\boldsymbol{Q}(t)$. To quantify the influence of the vibrating lattice on the state of the system we define the transfer efficiency as the highest depletion of the initial state for a given frequency $\omega$ and amplitude $A$. In Fig. \ref{resonanceFig}a (solid line) this transfer efficiency is presented as a function of frequency $\omega$ for amplitude $A=4$. It is clear that for two characteristic frequencies $\omega_1$ and $\omega_2$, the initial state can be totally depleted. Fig. \ref{resonanceFig}b presents the time dependence of occupations for the corresponding two frequencies. The full width at half maximum for both resonances is about $\delta \omega/(2\pi) \approx 700\,\mathrm{Hz}$. The characteristic frequencies almost do not depend on the amplitude $A$ and they are approximately equal to the energy difference between the appropriate eigenstates of the matrix $\hat{H}(\boldsymbol{Q}_0)$. Let me note that a full transfer of interacting atoms is obtained in a few milliseconds and therefore it is much faster than the experimentally obtained decay time of hundreds of ms \cite{Bloch}.

To show that the predictions described are almost insensitive to the approximations in the model, two additional tests were performed. Firstly, the expansion of the field operator $\Psi(\boldsymbol{r})$ were generalized so that all $d$-orbital states and their possible interactions were taken into account. The resulting transfer efficiency is shown with a dashed line in Fig. \ref{resonanceFig}a. As is seen, the previously predicted frequencies are slightly shifted and two additional peaks have appeared. They correspond to the resonant frequencies in which one of the interacting atoms is promoted to the $d_x$ or $d_y$-bands, respectively. This result shows that the described mechanism for creating higher orbital states is highly selective, and that for a particular choice of the resonant frequency one can neglect the other orbital states, i.e. the whole dynamics effectively takes place in a two dimensional subspace of coupled states. Secondly, to check the influence of the tunneling processes, the full many-body dynamics in a 1D optical lattice was studied. Due to  computational complexity, calculations were performed for eight atoms in a lattice with four sites, and with periodic boundary conditions. Vibrations $\boldsymbol{Q}=(32+4\sin(\omega t),20,8)$ with a frequency close to $\omega_2$ were considered. The simulations performed evidently show that for the chosen lattice parameters, the dynamics of the system can be treated as a dynamics carried out independently in each lattice site. Fig. \ref{resonanceFig}c shows the resulting transfer efficiency (filled circles) compared to single-site predictions (solid line) as well as the variance of the on-site number operator for the resonant frequency $\omega_2$ (inset). The situation changes significantly for larger tunneling amplitudes, i.e. for shallow lattices (about $12E_R$ for the example studied), and it will be discussed elsewhere. 

Additionally, let me briefly discuss the appreciably more complicated situation when the initial lattice is symmetric in both directions. Then, both $p$-band basis states have the same energy, but due to the existence of the coupling $U_{xy}$ they are not eigenstates of the Hamiltonian. As previously, resonant frequencies are determined by the eigenenergies of the Hamiltonian \eqref{HamMatrix} but now particles can be excited independently to symmetric or antisymmetric combinations of the original states $|\pm\rangle = (|020\rangle\pm |002\rangle)/\sqrt{2}$ by applying symmetric or antisymmetric lattice vibrations $\delta\boldsymbol{Q}_\pm(t) = (A,\pm A,0)\sin(\omega t)$, respectively. 

\begin{figure}
\includegraphics{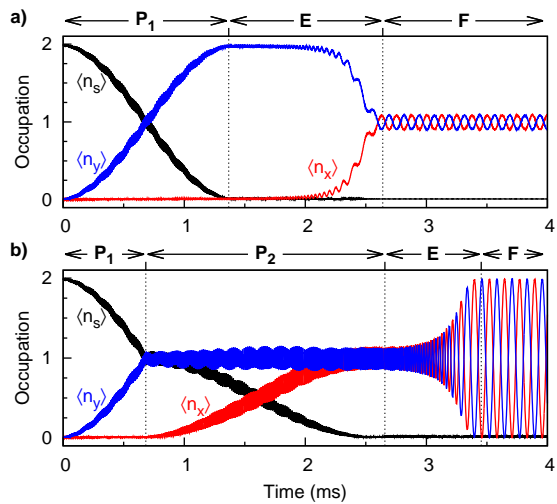}
\caption{Two experimental scenarios for the creation of a two particle superposition in excited band states. Initially the system is prepared in the ground state of a highly non symmetric lattice $\mathbf{Q}_0=(32,20,8)$. Plot (a): In the first scenario, two particles are transfered to the $p_y$ state by applying appropriate vibrations ($\mathbf{P}_1$) and then lattice depths in both directions are equilibrated ($\mathbf{E}$). If this process is slow enough, then the final state of the system is in the superposition $(|020\rangle -|002\rangle)/\sqrt{2}$. Plot (b): In the second case, two particles are transfered to the $p_y$ state by applying appropriate vibrations ($\mathbf{P}_1$). When the initial state is half-depleted the vibration frequency is changed to the other resonance, and the $p_x$ state is filled ($\mathbf{P}_2$). Then similarly to the previous scenario, lattice depths are equilibrated ($\mathbf{E}$). The final state is in a complex superposition of basis states and the occupation of each basis state varies in time ($\mathbf{F}$). At the moments when the occupation of $p_x$ and $p_y$ orbitals are equal, the system is in one of the vortex states $(|020\rangle \pm i|002\rangle)/\sqrt{2}$.}
\label{creationFig} 
\end{figure}

The mechanism discussed above can be easily extended to the idea of creating an orbital superposition of states $(|020\rangle + \mathrm{e}^{i\phi}|002\rangle)/\sqrt{2}$ with an arbitrarily chosen phase $\phi$. This can be done in a variety of different ways. Let me consider two scenarios which are efficient for creating states with $\phi=k\pi/2$, where $k=0,\ldots,3$ (see Fig. \ref{creationFig}). In both scenarios one starts with the system prepared in the static, non symmetric lattice with parameters $\boldsymbol{Q}_0=(32,20,8)$ in the insulating ground state with two particles in each lattice site. Then, the system is driven by a vibrating lattice with frequency $\omega_1$ ($\omega_2$) and the state $|002\rangle$ ($|020\rangle$) becomes occupied. In the first scenario (Fig. \ref{creationFig}a) one completly depletes the ground state (time interval $\mathbf{P_1}$). In the second one (Fig. \ref{creationFig}b) the driving frequency is switched to $\omega_2$ when the initial state has become half-depleted. Vibrations with frequency $\omega_2$ lead to the occupation of the state $|002\rangle$ ($\mathbf{P_2}$). In both scenarios, when the ground state becomes totally depleted, the lattice depth $q_x$ is brought down to equate $q_x$ and $q_y$ (interval $\mathbf{E}$). From this moment, the basis states $|020\rangle$ and $|002\rangle$ have the same energy, but due to the contact interactions (the last term in the Hamiltonian \eqref{NotResTerms}) they are not eigenstates of the Hamiltonian. The final state of the system ($\mathbf{F}$) is in some superposition of the basis states. The relative phase $\phi$ depends on the speed and details of the equilibrating process. Nevertheless, in the extreme case of adiabatic equilibration, the phase relationship between the orbital states is known.

In the first scenario at the beginning of the equilibration interval $\mathbf{E}$ the system is in an eigenstate (a ground or excited state in $p$-orbital subspace) of the initial Hamiltonian \eqref{HamMatrix}. Therefore, during an adiabatic equilibration of lattice parameters the system has to remain in that eigenstate of the temporal Hamiltonian. At the final moment, the states are $\frac{1}{\sqrt{2}}(|020\rangle\pm|002\rangle)$ (the relative phase is $0$ or $\pi$). A similar situation occurs in the second scenario. Before equilibration the system is in an almost equal superposition of states $|020\rangle$ and $\mathrm{e}^{i\Phi}|002\rangle$, which are the eigenstates of the Hamiltonian at that time. The phase $\Phi$ is very hard to control and therefore in practice it will be unknown. During the adiabatic equilibration the eigenstates track the time-evolving Hamiltonian, and change to the pair $(|020\rangle \pm |002\rangle)/\sqrt{2}$ in terms of basis states of the final Hamiltonian.  During this process the eigenstates accumulate phases $\Lambda_+$ and $\Lambda_-$, respectively. It can be shown straightforwardly that the final state of the system is $\mathrm{e}^{i\alpha}\left[\cos(\beta)|020\rangle + i \sin(\beta)|002\rangle\right]$, where $\alpha=(\Lambda_+ + \Lambda_- + \Phi)/2$ and  $\beta = \Lambda_+-\Lambda_--\Phi$. At later times (interval $\mathbf{F}$) the relative phase factor between $|020\rangle$ and $|002\rangle$ remains $\pm i$, while $\beta(t)$ evolves and leads to Rabi oscillations. At the moments when occupations of basis states are equal the state is in one of the vortex states $(|020\rangle \pm i|002\rangle)/\sqrt{2}$ (see Fig. \ref{creationFig}b).

Let me note that in the model discussed here, exitations to higher orbital states take place in all sites of the optical lattice not only independently but also simultaneously. Nevertheless, by applying an additional weak, anharmonic external potential one can make the energy gap between the orbitals site dependent, and, in consequence, the resonant frequencies can differ from site to site. This gives not only temporal but also spatial control of excitations. As such, it could be utilized for quantum computation engineering to address selected qubits formed by lattice sites with two bosons \cite{Deutsch}.

The results presented here show that lattice vibrations can effectively couple the ground Wannier state to chosen orbital states. The mechanism is very general since it originates from the fundamental Hubbard Hamiltonian by taking into account interactions between particles. In general, for each term describing a non vanishing interaction energy between particles in different single-particle states there always exist corresponding term describing a transfer of particles between them. In the static case these terms usually violate energy conservation and can therefore be neglected. However, when one considers time dependent Hamiltonians they should be always taken into account because they can significantly change the dynamics of the system. The additional terms in the Hamiltonian that were studied here should also be taken into account when rapid changes of the Hamiltonian are considered, e.g. to correctly describe quantum quench problems \cite{DeMarco}.

Finally, let me note that instead of driving the system via a changing depth of the lattice one can alternatively change the aspect ratio $\kappa$ or coupling constant $g$. The last possibility is quite interesting since it can be realized by applying an oscillating external magnetic field in the proximity of a Feshbach resonance. It means that by preparing appropriate pulses of the external magnetic field one could mimic interactions between particles and phonons by artificial vibrations propagating along the lattice (artificial phonons). Thus, the mechanism described here is not only an additional way of playing with orbital physics but it may also lead to a better understanding of those solid state problems that are beyond static theory \cite{Mermin}. 

The author thanks M. Brewczyk, P. Deuar, M. Gajda, B. Laburthe-Tolra, M. Lewenstein, and J. Mostowski for fruitful discussions. This research was funded by the National Science Center from grant No. DEC-2011/01/D/ST2/02019 and the EU STREP NAME-QUAM.

\end{document}